\documentclass[footinbib,aps,pra,reprint,superscriptaddress,nopacs]{revtex4-1}

\usepackage{graphicx}
\usepackage{amsmath,amsfonts,amssymb,bbm,bm}
\usepackage{verbatim}
\usepackage{braket}
\usepackage{mathtools}
\usepackage[absolute]{textpos}
\usepackage{ragged2e}

\usepackage{xcolor}

\DeclareMathOperator{\Tr}{Tr}
\DeclareMathOperator{\sF}{\mathcal{F}}
\DeclareMathOperator{\sP}{\mathcal{P}}
\DeclareMathOperator{\bx}{\mathbf{x}}

\begin{document}
\title{Fully Arbitrary Control of Frequency-Bin Qubits}

\author{Hsuan-Hao Lu}
\email{lu548@purdue.edu}
\affiliation{School of Electrical and Computer Engineering and Purdue Quantum Science and Engineering Institute, Purdue University, West Lafayette, Indiana 47907, USA}
\author{Emma M. Simmerman}
\affiliation{Quantum Information Science Group, Computational Sciences and Engineering Division, Oak Ridge National Laboratory, Oak Ridge, Tennessee 37831, USA}
\author{Pavel Lougovski}
\thanks{Present affiliation: Amazon Web Services, Seattle, Washington  98109, USA}
\affiliation{Quantum Information Science Group, Computational Sciences and Engineering Division, Oak Ridge National Laboratory, Oak Ridge, Tennessee 37831, USA}
\author{Andrew M. Weiner}
\affiliation{School of Electrical and Computer Engineering and Purdue Quantum Science and Engineering Institute, Purdue University, West Lafayette, Indiana 47907, USA}
\author{Joseph M. Lukens}
\email{lukensjm@ornl.gov}
\affiliation{Quantum Information Science Group, Computational Sciences and Engineering Division, Oak Ridge National Laboratory, Oak Ridge, Tennessee 37831, USA}
\date{\today}

\begin{abstract}
Accurate control of two-level systems is a longstanding problem in quantum mechanics. One such quantum system is the frequency-bin qubit: a single photon existing in superposition of two discrete frequency modes. %and a potential building block for scalable, fiber-compatible quantum information processing. 
In this work, we demonstrate fully arbitrary control of frequency-bin qubits in a quantum frequency processor for the first time. We numerically establish optimal settings for multiple configurations of electro-optic phase modulators and pulse shapers, experimentally confirming near-unity mode-transformation fidelity for all fundamental rotations. Performance at the single-photon level is validated through the rotation of a single frequency-bin qubit to 41 points spread over the entire Bloch sphere, as well as tracking of the state path followed by the output of a tunable frequency beamsplitter, with Bayesian tomography confirming state fidelities $\sF_\rho>0.98$ for all cases. Such high-fidelity transformations expand the practical potential of frequency encoding in quantum communications, offering exceptional precision and low noise in general qubit manipulation.
\end{abstract}

\maketitle
\textit{Introduction.---}The precise, coherent manipulation of the spectro-temporal properties of light has facilitated a plethora of applications, ranging from radio-frequency arbitrary waveform generation and optical communications~\cite{Cundiff2010,Torres2014}, to coherent control of chemical reactions~\cite{Teets1977,Silberberg2009} and extreme nonlinear optics~\cite{Baltuvska2003,Hassan2012}. %the ability to synthesize user-defined optical fields has left a permanent mark on photonic technologies, both inside and outside the laboratory.
Throughout these developments, Fourier-transform pulse shaping has played a central role, enabling arbitrary spectral filters that can shape optical fields on femtosecond timescales~\cite{Weiner2000, Weiner2011}. Concurrently, the Fourier dual process of electro-optic modulation has been a staple in fiber optics, functioning as ``temporal filters'' that multiply an input field in the time domain with phase patterns for applications such as optical communication~\cite{Kaminow2010} or frequency comb generation~\cite{Metcalf2013,Torres2014}. % and including data streams for communications~\cite{Kaminow2010}, high-frequency sinewaves for frequency comb generation~\cite{Metcalf2013,Torres2014}, and specialized analog patterns in waveform generation~\cite{Fontaine2010}.
The value of complex time-frequency control extends beyond classical optics to photonic quantum information processing (QIP) as well, with demonstrations of temporal shaping~\cite{Peer2005}, spectral coding~\cite{Lukens2014a}, wavepacket modulation~\cite{Kolchin2008,Liu2014}, spread spectrum~\cite{Belthangady2010}, electro-optic time lensing~\cite{Karpinski2017}, and high-dimensional quantum state reconstruction~\cite{Bernhard2013,Kues2017,Imany2018a} on single photons and entangled photon pairs. 

These successes have inspired the development of a complete QIP paradigm based on pulse shapers, modulators, and frequency-bin encoding~\cite{Lukens2017}. Drawing on arguments from linear-optical quantum computation (LOQC)~\cite{Knill2001}, the ``quantum frequency processor'' (QFP) approach has been shown scalable in principle, and a collection of gates comprising a universal set have been realized experimentally~\cite{Lu2018a,Lu2018b,Lu2019,IEEEptl2019}. Such scaling arguments prove crucial in establishing ultimate feasibility, yet leave many smaller---though highly practical---questions unanswered. For example, the fully arbitrary rotation of a single qubit represents a fundamental capability for any two-level system, enjoying a long history as a textbook example in quantum mechanics~\cite{Slichter1990, Mandel1995}. Yet neither a Solovay--Kitaev construction~\cite{Dawson2005} in terms of basic gates, nor general resource bounds, reveals the optimal construction of general two-mode unitaries, particularly when subject to practical resource constraints.

In this Letter, we answer this important question through theoretical analysis and experimental verification of arbitrary single-qubit gates in frequency-bin encoding. Our numerical simulations obtain three-element QFP configurations capable of any unitary operation with fidelity $\sF_W\geq0.9999$ utilizing single-tone modulation only; by either adding a second harmonic or cascading an additional pulse shaper/modulator pair, such operations achieve success probabilities $\sP_W>0.95$ or $\sP_W>0.999$, respectively. We reinforce these findings experimentally, synthesizing frequency-bin unitaries with performance in close agreement with theory. Finally, we highlight their use at the single-photon level in the rotation of a fixed input to arbitrary points on the Bloch sphere, obtaining output state fidelities $\sF_\rho>0.98$ with respect to the ideal. Our results represent the first full tomography of arbitrarily rotated frequency-bin qubit states, establishing resource guidelines for future systems and providing tools for fundamental applications in communications and coherent control.

\begin{figure*}[t!]
\centering\includegraphics[width=\textwidth]{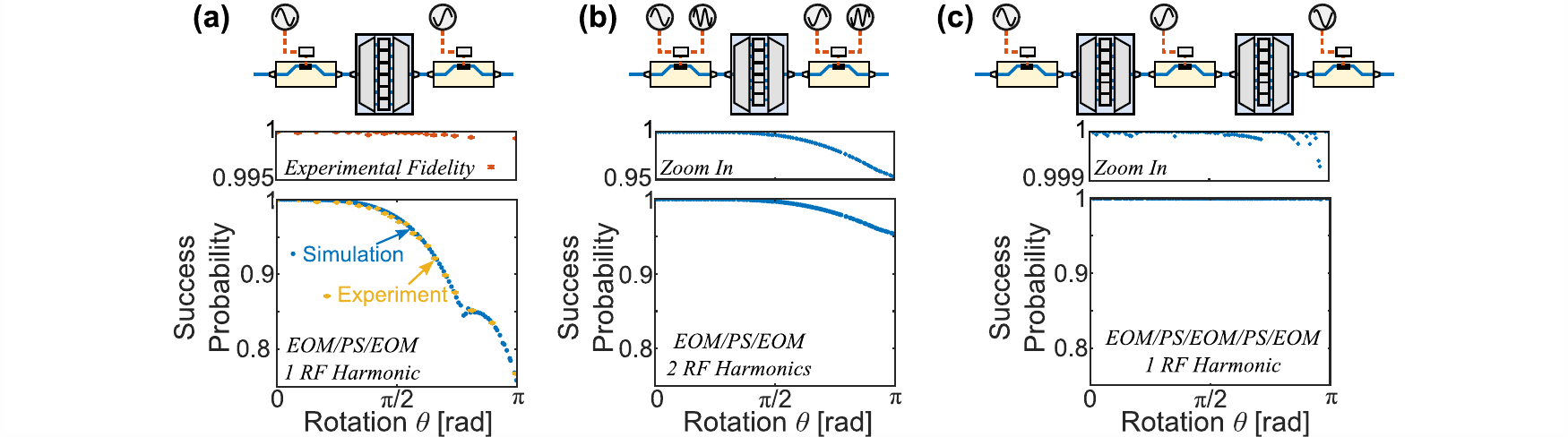}
\caption{Optimized success probability of single-qubit gate $U(\theta,0,0)$ for different QFP configurations. (a)~Three-element QFP, single-tone modulation (including experimental results). (See Appendix~\ref{appB} for discussion regarding the shoulder at $\theta \approx 3\pi/4$.) (b)~Three-element QFP, two-tone modulation. (c)~Five-element QFP, single-tone modulation. Zoomed-in plots for (b) and (c) detail the high-probability regions for each case. }
\label{fig1}
\end{figure*}

\textit{Problem Formulation.---}%In discrete-variable frequency-bin encoding, the fundamental unit of information is the qubit: 
A frequency-bin qubit can be represented as a single photon in a superposition of two frequency modes, or bins, described by annihilation (creation) operators $\hat{a}_0$ ($\hat{a}_0^\dagger$) and $\hat{a}_1$ ($\hat{a}_1^\dagger$) centered at frequency $\omega_0$ and $\omega_1$, respectively. A pure qubit state may be expressed as $\ket{\psi}=(c_0\hat{a}_0^\dagger + c_1\hat{a}_1^\dagger)\ket{\mathrm{vac}}$, where $\ket{\mathrm{vac}}$ is the vacuum state and $|c_0|^2 + |c_1|^2 = 1$. Logical basis states follow as $\ket{n}=\hat{a}_n^\dagger \ket{\mathrm{vac}}$ ($n\in\{0,1\})$. Any $2\times 2$ unitary operating on these modes can be parameterized as~\cite{Nielsen2000}
%\begin{widetext}
\begin{equation}
\label{e1}
U(\theta,\varphi,\lambda) = \begin{pmatrix}
\cos\frac{\theta}{2} & -e^{i\lambda}\sin\frac{\theta}{2} \\
e^{i\varphi}\sin\frac{\theta}{2} & e^{i(\varphi+\lambda)}\cos\frac{\theta}{2}
\end{pmatrix}, %
%= \begin{pmatrix}
%1 & 0 \\
%0 & e^{i\varphi}
%\end{pmatrix} \begin{pmatrix}
%\cos\frac{\theta}{2} & -\sin\frac{\theta}{2} \\
%\sin\frac{\theta}{2} & \cos\frac{\theta}{2}
%\end{pmatrix} \begin{pmatrix}
%1 & 0 \\
%0 & e^{i\lambda}
%\end{pmatrix} 
\end{equation}
%\end{widetext}
where $\theta\in[0,\pi]$, $\varphi\in[0,2\pi)$, and $\lambda\in[0,2\pi)$. %
Considering this as taking the inputs $\hat{a}_0$ and $\hat{a}_1$ to outputs $\hat{b}_0$ and $\hat{b}_1$, this implies that the output state coefficients, $\ket{\phi} = (d_0\hat{b}_0^\dagger + d_1\hat{b}_1^\dagger)\ket{\mathrm{vac}}$, satisfy $\begin{psmallmatrix} d_0 \\ d_1 \end{psmallmatrix} = U \begin{psmallmatrix} c_0 \\ c_1 \end{psmallmatrix}$. 

This mathematical formulation applies generally to any qubit system. The nuances of the QFP approach appear, though, when describing (i) the bins $\omega_0$ and $\omega_1$ as embedded within a comb spaced at $\Delta\omega$ ($\omega_n=\omega_0+n\Delta\omega; n\in\mathbb{Z}$) and (ii) the transformation on all modes $\hat{b}_m = \sum_n V_{mn}\hat{a}_n$ as characterized by an alternating series of electro-optic phase modulators (EOMs) driven with $\frac{2\pi}{\Delta\omega}$-periodic waveforms and pulse shapers applying arbitrary phases to each bin. %These two conditions---
%Temporal periodicity and line-by-line shaping ensure that the entire operation can be expressed as the discrete matrix $V$.
As modeled, $V$ is unitary over the entire countably infinite collection of bins%, i.e., $\sum_{m=-\infty}^{\infty} V_{mk}^* V_{mn} = \delta_{kn}$. Significantly
, though the $2\times 2$ submatrix in the computational space---call this $W=\begin{psmallmatrix} V_{00} & V_{01} \\ V_{10} & V_{11} \end{psmallmatrix}$---may or may not prove unitary, due to coupling into adjacent bins. 
  
While an apparent disadvantage of the QFP in this case (particularly when compared to the isolated modes of alternative frequency-bin approaches~\cite{Raymer2010,McGuinness2010, Kobayashi2016,Clemmen2016,Zhang2019b, Joshi2020}), this natural coupling between many bins %represents an valuable resource when considering more general QIP,  
%First, they can operate on multiple qubits concurrently, so that their uniform spacing permits 
facilitates multiphoton interference between all underlying modes as required for LOQC. Moreover, by cascading additional pulse shapers and EOMs and employing more complex modulation patterns, such adjacent-bin coupling can be fully compensated for, to realize smaller-dimensional gates with unity efficiency~\cite{Lukens2017,Lukens2020a}.

Within this overall framework, considerable progress has been made on a subset of $U(\theta,\varphi,\lambda)$: the phase-only gate $U(0,\varphi,\lambda)$ and the Hadamard $H=U(\frac{\pi}{2},0,\pi)$, with the former requiring only a single pulse shaper, and the latter realizable with an EOM/pulse shaper/EOM QFP~\cite{Lu2018a, Lu2018b}. These considerations engender optimism for experimental realization of arbitrary $U$, yet they do not answer the practical questions of explicit construction, nor elucidate the procedures involved in reconfiguring a given QFP for all possible unitaries.

\textit{Numerical simulations.---}For our simulations, we focus on pure-sinewave electro-optic modulation (either one or two tones), and QFPs with three or five elements. Limitation to odd-numbered QFPs follows from previous observations that adding a pulse shaper on either side of a QFP improves neither fidelity nor success probability, for any target operation. In fact, these remarks can be made rigorous in the present case of a single-qubit operation. Suppose that a particular QFP configuration realizes the gate $W=gU(\theta,0,0)$ (unitary up to an overall constant). Then, as derived in Appendix~\ref{appA}, the same QFP can actualize the gate $gU(\theta,\varphi,\lambda)$ by delaying the rf signals applied to the first and last EOM and adding linear phases to the first and last pulse shaper (or the single pulse shaper in a three-element QFP).

This finding implies that, for the purpose of establishing performance under system constraints, one need only concentrate on $U(\theta,0,0)$ numerically. We emphasize that, while similar, these phase degeneracies prove fundamentally more significant than those resulting from the freedom to set a phase reference. As argued in Ref.~\cite{Rahimi2013} and invoked below in our own characterization procedure, the prerogative to define the ``in-phase'' condition across modes at the input and output planes of an optical multiport simplifies the process of extracting $V_{mn}$. %amplitudes and phases of the individual transformation elements $V_{mn}$. 
However, such phase reference flexibility does not imply the physical equivalence of operations that differ by this reference. For example, if one defines the reference so that the QFP realizes $U(\theta,0,0)$, modifying the transformation to $U(\theta,\varphi,\lambda)$ produces measurable differences in the output state, impacting any subsequent operations downstream. Accordingly, the relationship between phase and EOM delay discussed here is not just the establishment of a reference: it gives a means to realize a $(\varphi,\lambda)$ combination for \emph{any} reference definition.

To benchmark the performance of single-qubit gates synthesized on the QFP, we randomly generate 150 samples of $\theta\in[0,\pi]$, %[corresponding to 150 different unitaries $U(\theta,0,0)$], 
and numerically find the solutions $U(\theta,0,0)$ for three different scenarios (see Fig.~\ref{fig1}): three-element QFP driven by (a) one or (b) two rf tones and (c) five-element QFP driven by single tone. Case~(a) is the baseline QFP which we have utilized in previous experiments~\cite{Lu2018a,Lu2018b,Lu2019}, while cases~(b) and (c) describe the two most immediate upgrades; (b) has been explored in a limited context for a frequency-bin qutrit operation~\cite{Lu2018a}, while (c) has so far required too many resources for implementation. Yet all three are realizable with standard, commercially available components. %---no specialized rf or optical equipment required.
We then assess the performance of $W$ with respect to the desired $U$ according to gate success $\sP_W = \Tr (W^\dagger W)/2$ and fidelity $\sF_W = |\Tr (W^\dagger U)|^2 /(4\sP_W)$ metrics, where $\sP_W$ describes the probability of a photon remaining in the computational space, and $\sF_W$ defines the quality of the operation~\cite{Uskov2009}. Our goal is to maximize $\sP_W$ while constraining $\sF_W\geq 0.9999$~(Appendix~\ref{appB}).

Figure~\ref{fig1} plots the simulation results. $\sP_W$ shows a strong dependence on $\theta$, suggesting that those unitaries with small $\theta$ are easier to realize. This matches our intuition as the identity and phase-only gates ($\theta=0$) can be realized without any EOM, while gates like Pauli $X$ and $Y$ ($\theta=\pi$) require proper engineering of the mixing process such that the photon can be completely hopped to the opposite bin. Additionally, the results indicate that gate performance can be significantly boosted %with increased complexity %in the controls,
by introducing either an additional rf harmonic [Fig.~\ref{fig1}(b), $\sP_W > 0.95$] or extra components [Fig.~\ref{fig1}(c), $\sP_W > 0.999$]. %, which comes as no surprise that the increased complexity can lead to more complex operations. 
%For example, our previous experimental demonstration~\cite{Lu2018a} has shown that a three-element QFP with two rf tones can synthesize a three-dimensional DFT gate---the $3\times 3$ extension of the Hadamard. 
Here, we experimentally focus on the setup in Fig.~\ref{fig1}(a) due to equipment availability---i.e., insufficient rf amplifier bandwidth for (b), lack of EOMs and pulse shapers for (c).
% Talk about "phase transition effect for Fig. 1(a)
% Talk about why these simulations are significant compared to our previous work
%Due to limited number of components to construct larger QFP circuit or generate the second-harmonic RF tone at the required power level, we elect to implement the first scenario in the experiments below.

\begin{figure}[!bt]
\includegraphics[width=3.4in]{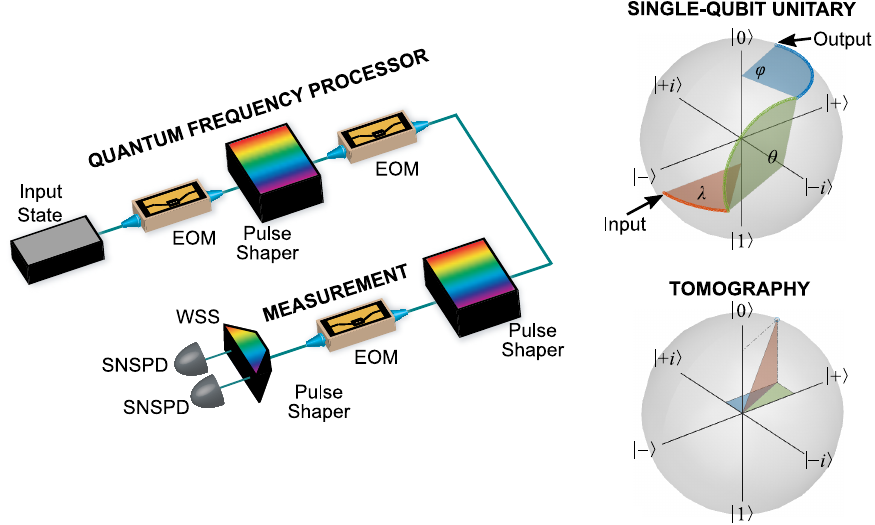}
\caption{Experimental setup for arbitrary frequency-qubit operations and state measurement. Insets show an example unitary rotation for an input state $\ket{\psi}=0.6\ket{0}-0.8\ket{1}$ and $(\theta,\varphi,\lambda)=(0.7\pi,0.55\pi,0.25\pi)$, with tomography represented in terms of projections onto each axis. Here $\ket{\pm}\propto\ket{0}\pm\ket{1}$ and $\ket{\pm i}\propto\ket{0}\pm i\ket{1}$.}
\label{fig2}
\end{figure}

\textit{Gate characterization.---}Figure~\ref{fig2} provides a schematic of the experimental setup. A high-frequency rf oscillator generates 25-GHz sinusoidal voltages to drive both EOMs~\cite{footnote1}, where their amplitudes and delays are set with manual phase shifters and attenuators. Meanwhile, the optimized spectral phase pattern is programmed onto the QFP shaper. Experimentally, we select 21 out of the 150 previous solutions from Fig.~\ref{fig1}(a). % previously obtained from the numerical optimizer for implementation.
To investigate whether each gate performs as anticipated, we utilize a coherent-state--based characterization approach~\cite{Rahimi2013,Lu2018a} by probing our QFP with an electro-optic frequency comb. %and measuring the output spectrum for different input superpositions. 
As a result, we are able to reconstruct the mode-transformation matrix $W$ and compute the experimental $\sF_W$ and $\sP_W$, as shown in Fig.~\ref{fig1}(a). All measured gate fidelities are above $0.9993$ (except for one, unexplained outlier), and the success probabilities track closely the theoretical prediction.%, highlighting the fine controllability of our operation. 

\textit{Arbitrary state rotation.---}The previous tests confirm synthesis of arbitrary frequency-qubit operations, when viewed in terms of optical modes. Yet in the context of photonic QIP, these \emph{mode} transformations are valuable insofar as they enable high-fidelity operations on quantum \emph{states}.  %(single photons in discrete-variable encoding).
Accordingly, we explore these gates at the single-photon level, focusing specifically on their ability to convert a fixed input to an arbitrary output state. %For example (see Fig.~\ref{fig2}), given a well-defined qubit $|\Psi\rangle$ as the input of the QFP, we can convert it to an output state anywhere on the Bloch sphere through an unitary operation $U(\theta,\varphi,\lambda)$.
We can then assess the quality of this manipulation by performing quantum state tomography (QST) on the output photon. 

Following the QFP with a set of projective measurements, we reconstruct its density matrix ($\hat{\rho}$) through QST and compute the state fidelity with respect to the ideal output state $\ket{\phi}$ via $\sF_{\rho}=\braket{\phi|\hat{\rho}|\phi}$. (Note the change in definition from the Hilbert--Schmidt fidelity $\sF_W$ utilized for matrix characterization.) We prepare a single-photon-level source by attenuating a continuous-wave laser at frequency $\omega_0$ to $\sim$10$^6$~counts/s (1/10 of the detector saturation level) prior to the QFP. Since neither the QFP operation nor QST involve multiphoton interference, the results of a weak coherent state are fully equivalent to those of true single photons at the same average flux. To show that we can bring this input state, $\ket{0}=\hat{a}_0^\dagger\ket{\mathrm{vac}}$, at the north pole of the Bloch sphere, to any arbitrary state within the whole sphere, we choose 11 values of $\theta\in[0,\pi]$ and assign a few different $\varphi$ to each, amounting to a total of 41 gates to implement. The ideal output state is
$\ket{\phi}=\cos\frac{\theta}{2}\ket{0}+e^{i\varphi}\sin\frac{\theta}{2}\ket{1}$.
%, where, for the sake of convenience, we set $\lambda=0$ since it has no effect on the input state $|0\rangle$. 

For single-qubit QST~(Appendix~\ref{appC}), we perform three Pauli measurements ($Z$, $X$, and $Y$) to project the output state onto the eigenvectors $\ket{t}$ (six in total): $\{\ket{0},\ket{1},\ket{\pm},\ket{\pm i} \}$. Measuring in $Z$, $X$, and $Y$ is equivalent to applying $\mathbbm{1}$, $H$, and $HS^\dagger$ prior to computational-basis projection, where $S=\begin{psmallmatrix} 1 & 0 \\ 0 & i \end{psmallmatrix}$. $Z$ measurement demultiplexes the photons by color with a wavelength-selective switch (WSS), and records the counts in $|0\rangle$ and $|1\rangle$ with superconducting nanowire detectors (SNSPDs). %The $H$ required for tomography can be realized efficiently with another three-element QFP~\cite{Lu2018a,Lu2018b}, but due to component availability 
For the $H$ required for tomography, we implement a simpler probabilistic Hadamard gate based on a single EOM, with a sinusoidal rf voltage chosen for equal mixing probability between $\omega_0$ and $\omega_1$~(see Appendix~\ref{appC} and Ref.~\cite{ImanyHOM} for more details).  %The relatively high scattering into adjacent bins (40\% for this single EOM mixer) reduces the efficiency of the tomographic process but does not impact projection fidelity. 
We precede this EOM with another pulse shaper to apply the $S^\dagger$ gate and block any residual photons outside of the single-qubit space after the QFP. For each measurement setting, we record the counts over 1~s, then subtract the average detector dark counts and obtain a final dataset $\mathcal{D}=\{N_0,N_1,N_+,N_-,N_{+i},N_{-i} \}$ with all outcomes for subsequent tomographic analyses.

\begin{figure}[b!]
\includegraphics[width=3.4in]{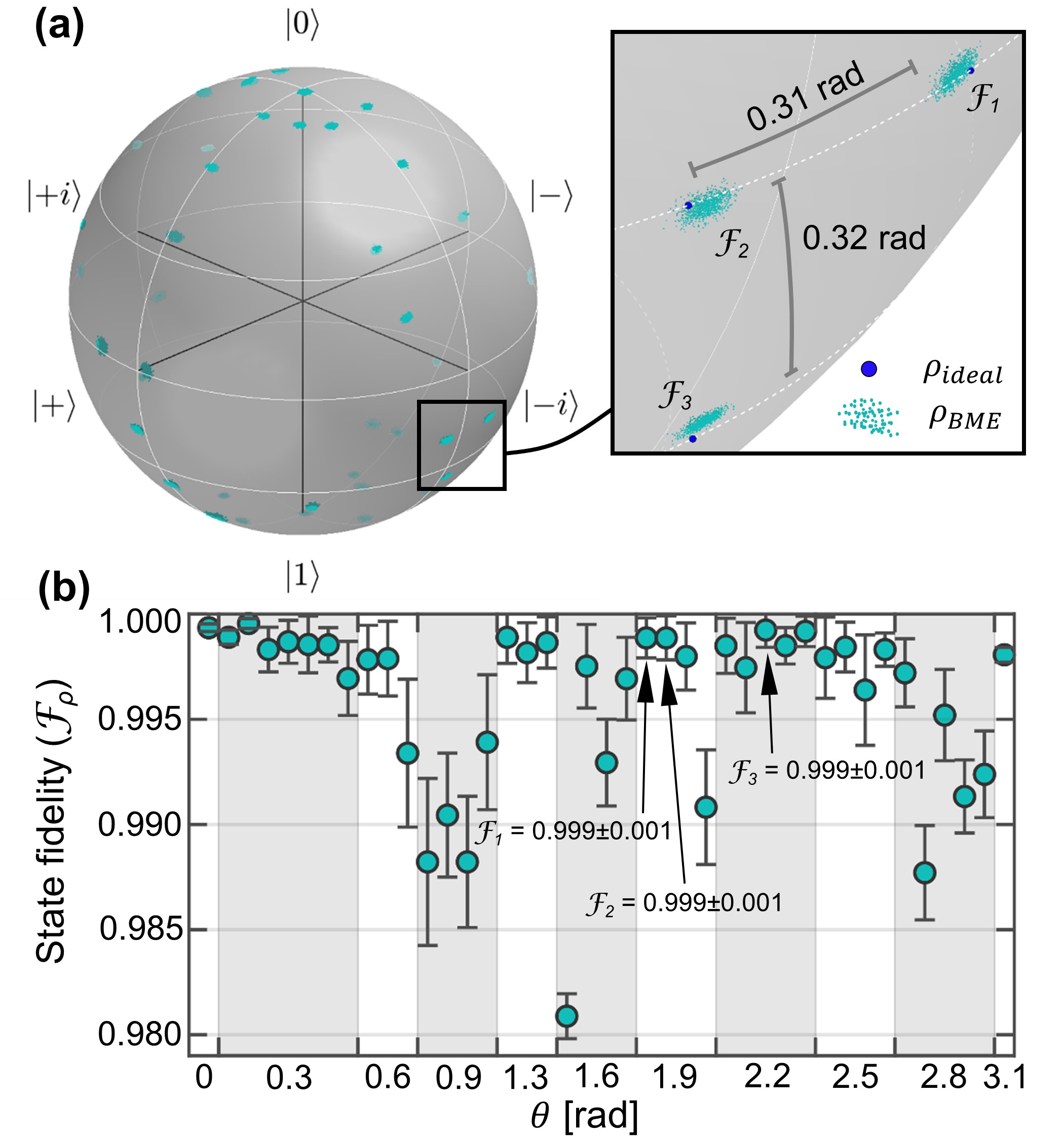}
\caption{Arbitrary single-qubit rotations on the QFP. (a) Retrieved Bayesian samples plotted on the Bloch sphere (green dots) following 41 different transformations $U(\theta,\varphi,0)$. Three examples are highlighted in the inset, where blue dots mark the corresponding ideal states. (b) Associated state fidelities, grouped by $\theta$ value, with each data point corresponding to a randomly chosen $\varphi$. The mean and standard deviation are computed from $1024$ Bayesian samples.}
\label{fig3}
\end{figure}

For reconstruction, we employ Bayesian mean estimation~\cite{Blume2010}, an advanced tomographic technique which avoids unjustifiably low-rank estimates and furnishes natural error bars. %With each count dataset $\mathcal{D}=\{N_0,N_1,N_+,N_-,N_{+i},N_{-i} \}$, 
We parameterize the density matrix $\hat{\rho}(\bx)$ and sample a posterior distribution $\pi(\bx)\propto L_\mathcal{D}(\bx) \pi_0(\bx)$ with multinomial likelihood
%\begin{equation}
%\label{likelihood}
$L_\mathcal{D}(\bx) = p_0^{N_0} p_1^{N_1} p_+^{N_+} p_-^{N_-} p_{+i}^{N_{+i}} p_{-i}^{N_{-i}},$
%\end{equation}
where $p_t \equiv \braket{t|\hat{\rho}(\bx)|t}$ is the probability of measuring the state $\ket{t}$ given the proposed state $\hat{\rho}(\bx)$. We adopt the parametrization, prior distribution $\pi_0(\bx)$, and sampling procedure recently proposed in Ref.~\cite{Lukens2020b}, obtaining $R=1024$ density matrix samples $\hat{\rho}_r$ for each tomographic dataset, from which we estimate the fidelity according to the mean and standard deviation of the values of the individual samples ($\sF_r = \braket{\phi|\hat{\rho}_r|\phi}$).
%mean $\overline{\sF}_\rho=\frac{1}{R}\sum_{r=1}^R  \braket{\phi| \hat{\rho}_r |\phi}$ and standard deviation $\Delta\sF_\rho = \left[ \frac{1}{R}\sum_{r=1}^R  (\braket{\phi| \hat{\rho}_r|\phi})^2 - \overline{\sF}_\rho^2\right]^{1/2}$.

Figure~\ref{fig3} depicts the QST results. We map the ideal output states and the retrieved Bayesian samples onto the Bloch sphere [Fig.~\ref{fig3}(a)]. Three of the transformations are highlighted in the zoomed-in inset, where the Bayesian samples follow the ideal states closely. This suggests strong agreement between the design and experimental implementation, confirmed by Bayesian mean state fidelities above 98\% across all gates [Fig.~\ref{fig3}(b)].

\begin{figure}[b!]
\includegraphics[width=3.4in]{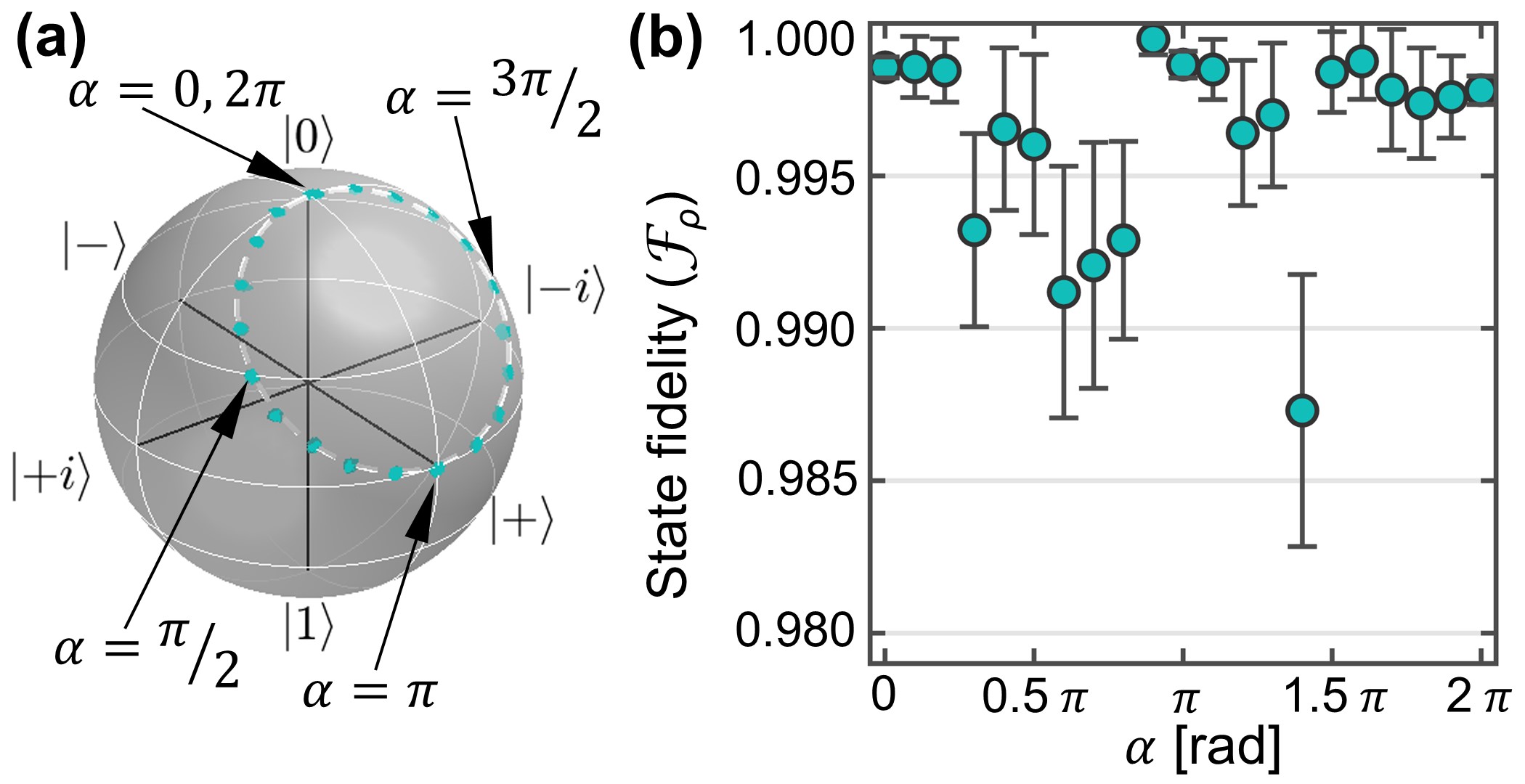}
\caption{Tunable beamsplitter. (a) Ideal output state trajectory (dashed line) and retrieved Bayesian samples (green dots) plotted on the Bloch sphere. (b) Bayesian state fidelities as function of pulse shaper phase $\alpha$.}
\label{fig4}
\end{figure}

\textit{Tunable beamsplitter.---}In addition to a randomly chosen set of single-qubit rotations, we can also explore coherent quantum state control across a specified trajectory. %on the QFP by properly setting the phase patterns on EOMs and pulse shaper to match those obtained from the optimized solutions. The versatility of QFP, however, can go beyond that.
Previously, we found a set of analytical solutions for tunable frequency beamsplitters~\cite{Lu2018b}, where the reflectivity can be set anywhere between 0 and $0.5$ simply by changing the depth of the phase shift $\alpha$ imparted by the QFP shaper between frequency bins 0 and 1 (while both EOMs remain fixed; see Appendix~\ref{appD} for more details). We sample 21 evenly spaced $\alpha\in[0,2\pi]$ for implementation, and again repeat the QST measurement for the same state input $\ket{0}$. Figure~\ref{fig4} depicts the experimental results. As we increase $\alpha$, the output state is moved from the north pole ($\ket{0}$) to the equator ($\ket{+} = H\ket{0}$), and then back to the north pole ($\ket{0}$), following a counterclockwise trajectory on the Bloch sphere (dashed line in Fig.~\ref{fig4}). Again all measurements are in excellent agreement with theory ($\mathcal{F}_\rho > 0.98$). %as the state is coherently moved around the sphere surface.

\textit{Discussion.---}In addition to addressing fundamental questions in frequency-bin quantum state control, the findings described here appear particularly relevant in the applications of quantum communications and networking. Indeed, one of the inherent benefits of frequency-bin encoding is its compatibility with fiber-optic communications: %discrete frequency channels are well-matched to single-mode fiber,
the QFP paradigm already leverages commercial telecom components (EOMs and pulse shapers), and frequency-bin operations can be extensively parallelized according to the principles of wavelength-division multiplexing (WDM). This synergy has enabled several recent quantum networking demonstrations invoking WDM for distributing entanglement in other degrees of freedom~\cite{Lim2008,Aktas2016,Wengerowsky2018,Lingaraju2020}; the QFP approach moves even further by exploiting frequency bins for encoding quantum information as well, and the fully arbitrary unitaries realized here should make feasible an array of quantum networking protocols with frequency bins. Whereas the previously shown Hadamard ($\theta=\frac{\pi}{2}$)~\cite{Lu2018a,Lu2018b} would be sufficient (along with the identity) for basis measurements in quantum key distribution~\cite{Gisin2002}, it is only through these more general unitaries that the full range of qubit quantum information protocols can be realized. For example, both superdense coding~\cite{Bennett1992} and quantum teleportation~\cite{Bennett1993} require single-qubit gates including a full 180$^\circ$ rotation ($\theta=\pi$), and the standard CHSH Bell inequality~\cite{Clauser1969} relies on measurements preceded by unitaries with $\theta\in\{0,\frac{\pi}{4},\frac{\pi}{2},\frac{3\pi}{4}\}$.

Moreover, while we have focused specifically on the fundamental two-level qubit here, one of the salient features of the frequency degree of freedom is its natural compatibility with high-dimensional qudit ($d>2$) encoding~\cite{Imany2018a,Kues2017,Ikuta2019,Erhard2020}. Importantly, the same design procedure adopted here can be applied for the construction of arbitrary qudit operations as well. %As dimensionality increases, though, so do the required experimental resources.
As initial examples, we have numerically found $2d+1$ EOMs and pulse shapers sufficient for high-dimensional frequency hopping (up to $d=5$) using single-tone rf modulation~\cite{Lukens2020a}; we have also found $d-1$ rf harmonics capable of realizing $d$-dimensional discrete Fourier transformations (up to $d=10$) on a single three-element QFP. The main limitations moving to higher dimensions, then, are technical in nature---namely, the complexity of rf drive waveforms and the number of pulse shapers and EOMs available.
%~\cite{footnote2}.

On the characterization side, our focus on QST of an arbitrarily rotated state corroborates the gate performance estimated from classical measurements. On the other hand, quantum process tomography (QPT) would be required for a complete quantum-mechanical description of the gate itself~\cite{Chuang1997}. This procedure relies on preparation of multiple input states (four in the case of a single-qubit operation), followed by QST of each output after the QFP, which would necessitate additional components beyond those available to us. %an additional EOM and pulse shaper for state preparation beyond those available to us in the lab. 
Given our understanding of the physical mechanisms involved in the QFP, %as well as good agreement between our Bayesian QST measurements and those expected from an underlying operation consisting of a mixture of the ideal unitary and pure white noise~\cite{Supplemental},
we do not expect fundamentally new insights from QPT. Nevertheless, realization of complete QPT---perhaps leveraging Bayesian techniques for experimental simplifications---would prove valuable in future work, as a means to further validate performance.

\begin{acknowledgments}
This research was performed in part at Oak Ridge National Laboratory, managed by UT-Battelle, LLC, for the U.S. Department of Energy under contract no. DE-AC05-00OR22725. Funding was provided  by the U.S. Department of Energy, Office of Science (Office of Advanced  Scientific Computing Research, Early Career Research Program; and Office of Workforce Development for Teachers and Scientists Science Undergraduate Laboratory Internship Program) and the National Science Foundation (Grant No. 1839191-ECCS).
\end{acknowledgments}

\appendix
\section{Single-Qubit QFP Transformation Symmetries}
\label{appA}
Consider a QFP composed of $N+1$ EOMs and $N$ pulse shapers in an alternating series, and configured to realize single-qubit transformation $W=gU(\theta,0,0)$ [see Eq.~(1) in the main text for the definition], where $|g|^2\leq 1$ represents the gate success. Assume the first and last EOMs are driven by $A(t)$ and $B(t)$ (each $\frac{2\pi}{\Delta\omega}$-periodic waveforms), and the first and last pulse shapers are programmed with spectral phases $p_k$ and $q_k$ on the $k$-th frequency mode. The corresponding transformation by frequency multiport $V$, with projection of $W$ onto the single-qubit space ($m,n\in\{0,1\}$) is then
\begin{equation}
\label{eq1}
\tag{S1}
W_{mn} = \sum_{k=-\infty}^{\infty} \sum_{l=-\infty}^{\infty} d_{m-k}e^{iq_k}T_{kl}e^{ip_l}c_{l-n}
\end{equation}
where $T_{kl}$ is the mode transformation from all elements in the QFP apart from the first and last EOM/shaper pair. In the case of $N=1$ (EOM/shaper/EOM QFP), $T$ is an identity matrix and the center pulse shaper is programmed with $p_k+q_k$. The factors $c_{l-n}$ and $d_{m-k}$ are the mode coupling coefficients between modes $n$ and $l$ and modes $k$ and $m$, for the first and last EOM, respectively. They represent the Fourier series coefficients of the periodic modulation, and can be expressed as
\begin{equation}
\tag{S2}
\begin{aligned}
\label{Fourier}
c_{l-n}=\frac{1}{T}\int_T dt\, e^{iA(t)} e^{i(l-n)\Delta\omega t}\\
d_{m-k}=\frac{1}{T}\int_T dt\, e^{iB(t)} e^{i(m-k)\Delta\omega t},
\end{aligned}
\end{equation}
where the integration is over any full period $T=\frac{2\pi}{\Delta\omega}$.

To actualize $U(\theta,\varphi,\lambda)$, our goal is to reconfigure the QFP such that the new mode transformation $\tilde{W}_{mn}$ equals  $e^{i(m\varphi+n\lambda)}W_{mn}$, or $gU(\theta,\varphi,\lambda)$ by specification. Suppose that we delay the rf signals applied to the first and last EOM by $\tau_a$ and $\tau_b$, respectively, and introduce additional phase shifts $\delta_k$ and $\epsilon_k$ to the first and last pulse shaper, respectively. We obtain the modified mode transformation
\begin{equation}
\tag{S3}
\begin{split}
\label{e3}
\tilde{W}_{mn} &= e^{i\Delta\omega(m\tau_b-n\tau_a)} \sum_{k=-\infty}^{\infty} \sum_{l=-\infty}^{\infty} e^{i(\epsilon_k-k\Delta\omega\tau_b)}\\
& \times\left[d_{m-k}e^{iq_k}T_{kl}e^{ip_l}c_{l-n}\right]e^{i(\delta_l+l\Delta\omega\tau_a)}
\end{split}
\end{equation}
In order to fulfill $\tilde{W}_{mn} = e^{i(m\varphi+n\lambda)}W_{mn}$, we can set the delays such that $\Delta\omega\tau_b=\varphi$, $\Delta\omega\tau_a=-\lambda$, and make the double summation in Eq.~(\ref{e3}) identical to $W_{mn}$ by choosing $\epsilon_k=k\Delta\omega\tau_b$ and $\delta_l=-l\Delta\omega\tau_a$. Thus, we arrive at a simple method for reconfiguring the QFP for $U(\theta,\varphi,\lambda)$ given $U(\theta,0,0)$: delay the rf signals applied to the first and last EOM by $\tau_a=-\frac{\lambda}{\Delta\omega}$ and $\tau_b=\frac{\varphi}{\Delta\omega}$, respectively, and add linear phases $\delta_k=k\lambda$ and $\epsilon_k=k\varphi$ to the first and last pulse shaper, respectively. The remaining settings of the QFP are unchanged. This procedure readily extends to higher-dimensional unitaries. For example, if the QFP is originally programmed to implement $U$, we could follow the same method to reconfigure QFP and realize $D_1 U D_2$, as long as $D_1$ ($D_2$) is a diagonal unitary with a constant phase increment $\varphi$ ($\lambda$) across the diagonal elements. Intuitively, this process works because a linear phase is equivalent to a delay (while for a two-dimensional system, \emph{any} arbitrary phase shift between two modes can be seen as a delay); tuning the bookend EOMs redefines the input/output phase references, while the pulse shaper corrections ensure that, inside the QFP, the frequency-bin mixing operation proceeds unaffected.

\section{Numerical Optimization}
\label{appB}
\begin{figure}[!t]
\includegraphics[width=\columnwidth]{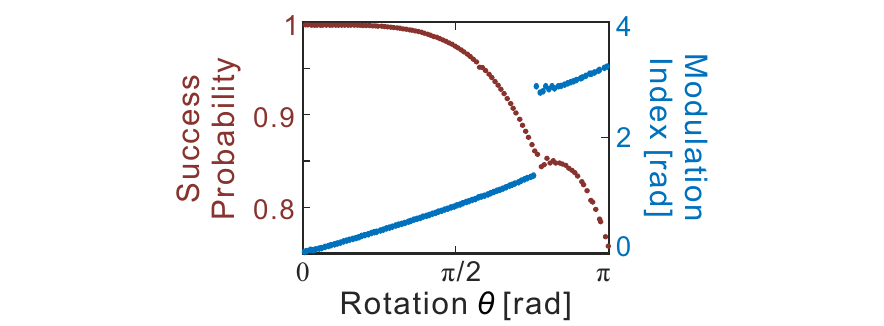}
\caption{Numerical simulation of single-qubit transformation $U(\theta,0,0)$ on a EOM/pulse shaper/EOM QFP driven by single RF tone, with fidelity constrained $\sF_W\geq0.9999$. Both the gate success probability and the EO modulation index show strong dependence on the parameter $\theta$ (rotation around the $y$-axis). In all solutions, we find both EOMs share the same modulation index.}
\label{figS1}
\end{figure}

In this section, we highlight some of the features found in the numerical optimization for single-qubit rotation using an EOM/shaper/EOM driven by a single rf tone. Given practical limitations on the attainable rf power level, we restrict the modulation index (i.e., peak phase shift) to less than 4~rad in all simulations. Across all the transformations we simulate, the temporal modulations on both EOMs in the optimized solution are always time-shifted replicas (i.e., same modulation index). Interestingly, this trend seems to hold even when we extend to more rf harmonics or a higher-dimensional system~\cite{Lu2018a}, but does not transfer to the case of larger QFP circuits (more components). We plot the modulation index---i.e., peak phase shift---with respect to $\theta$ in Figure~\ref{figS1} which, similar to the success probability, also shows a strong dependence on $\theta$. In our preliminary simulations utilizing a nonlinear constrained multivariate optimizer alone (\emph{fmincon} in MATLAB), we noticed that the optimizer tended to converge toward two different families of solutions---one with a smaller modulation index, and the other with larger modulation index---depending on the randomly assigned starting points. The former (latter) shows better success probability when $\theta$ is below (above) $0.76\pi$, which explains the gap and the shoulder in both curves in Figure~\ref{figS1}. As we notice that \emph{fmincon} is susceptible to local extrema trapping near the starting points, we have to rerun the optimizer multiple times with different initial points to ensure the optimal solution can be found.

This is one of the motivations for us to introduce a hybrid algorithm using particle swarm optimization (PSO)---a nature-inspired metaheuristic algorithm---together with \emph{fmincon} for better convergence. PSO starts with a number of particles moving around in the $M$-dimensional problem space ($M$ is the number of variables) in search of the extremum iteratively, where each particle represents a possible solution. In each iteration, the particle moves toward a new position depending on both the history of its local best known position and that of the entire swarm. The optimizer halts when all the particles converge to a single point in the problem space. A potential issue for traditional PSO is that we can no longer maximize one metric while constraining another, which in our case is the gate success probability ($\sP_W$) and fidelity ($\sF_W\geq0.9999$), respectively. Therefore, we follow the procedures proposed in Ref.~\cite{Parsopoulos2002} to lump both the gate success probability and fidelity into a single cost function. Specifically, the cost function we try to minimize is $C=-\sP_W+H(\sF_W)$, where $H(\sF_W)=\beta(\sF_W) \cdot (0.9999-\sF_W)$ and $\beta(\cdot)$
is a multistage, relative-violated function. We choose
\begin{equation}
\tag{S4}
\beta(\sF_W) = \begin{cases}
100 &;\;\;  0\leq \sF_W < 0.9\\
50 &;\;\;  0.9 \leq \sF_W < 0.99\\
25 &;\;\; 0.99 \leq \sF_W < 0.999\\
10&;\;\;  0.999 \leq \sF_W < 0.9999\\
0&;\;\;  0.9999 \leq \sF_W < 1,
\end{cases}
\end{equation}
which reduces the penalty for fidelity as it approaches unity.
%(for example, $\beta=100, 50, 25, 10$ when $\sF_W<0.9, 0.99, 0.999, 0.9999$).
For a sanity check, we follow the PSO with another quick run of \emph{fmincon} using the best solution found in PSO as the initial point and see if the performance can be further improved.

\section{Quantum State Tomography (QST): Experimental Methods}
\label{appC}
\begin{figure}[!b]
\includegraphics[width=\columnwidth]{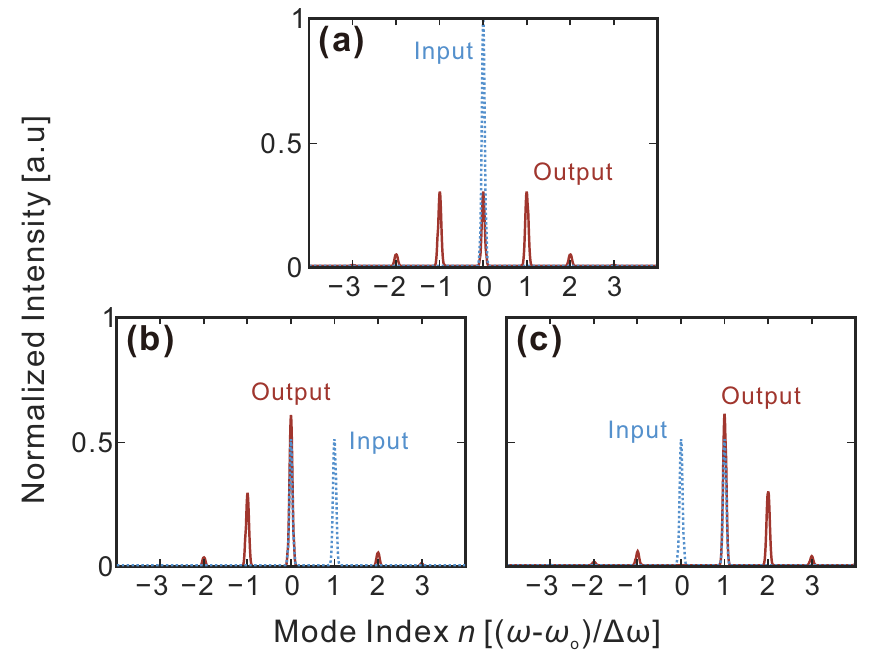}
\caption{Probabilistic Hadamard operation for QST. Example of simulated output spectra for specific inputs. (a) Pure mode~$0$, $\ket{0}$. (b) Mode~$0$ and~$1$ in phase, $\ket{+}\propto\ket{0}+\ket{1}$. (c) Mode~$0$ and~$1$ out of phase, $\ket{-}\propto\ket{0}-\ket{1}$. 40\% of the photons are scattered outside of the computational space (mode~$0$ and~$1$) due to the usage of single EOM.}
\label{figS2}
\end{figure}
\vspace{-0.1in}
As discussed in the main text, to realize three Pauli measurements ($Z$, $X$, and $Y$) for QST, one needs to apply $\mathbbm{1}$, $H$, and $HS^\dagger$ prior to computational-basis measurement, where $H$ is the Hadamard operation and $S=\begin{psmallmatrix} 1 & 0\\ 0 & i \end{psmallmatrix}$. Experimentally, we follow the gate operation (namely, the last EOM in the QFP) with a pulse shaper to filter out any photons outside of mode $0$ and $1$ such that residual scattering will not affect the operations downstream---i.e., by coupling back to the computational space and introducing measurement errors. For the $H$ operation, we elect to use a single EOM to realize a probabilistic version of the Hadamard gate; driving this EOM with a sinusoidal rf voltage with a modulation index of 1.434~rad, we have equal power splitting between mode $0$ and $1$, as shown in Fig.~\ref{figS2}(a). This approach inevitably scatters more photons ($\sim$40\%) out of the single-qubit space than its near-deterministic counterpart using a three-element QFP~\cite{Lu2018a}, but can be implemented with the limited number of EOMs available in our laboratory. Figure~\ref{figS2}(b) and (c) show examples of projecting $\ket{+}$ and $\ket{-}$ states to $\ket{0}$ and $\ket{1}$, respectively. The clear contrast between two output frequency modes shows that this approach, while reducing the efficiency of the tomographic process, does not sacrifice projection accuracy. 

We emphasize that the rf drive applied to this EOM is synchronized with those in the QFP, and its relative timing should be properly set as well. Experimentally, we first program a Hadamard operation on the QFP and send the output photons (now in the $\ket{+}$ state) through the pulse shaper (applying zero phase to all bins) and EOM for tomography, and fine tune the delay of this electrical drive via a manual rf phase shifter while recording the photon counts in two computational modes. Once we obtain maximal (minimal) amount of photons in the frequency mode $0$ (mode $1$), this EOM is aligned to perform the desired $H$ measurement and the delay setting is fixed throughout the rest of the experiment. Then, by applying a phase of $-\frac{\pi}{2}$ to frequency bin 1 on the measurement pulse shaper, we can realize the necessary $S^\dagger$ operation for $\ket{\pm i}$ measurement as well.

%\vspace{0.1in}
\section{Tunable Beamsplitter Design}
\label{appD}
\begin{figure}[!t]
\includegraphics[width=\columnwidth]{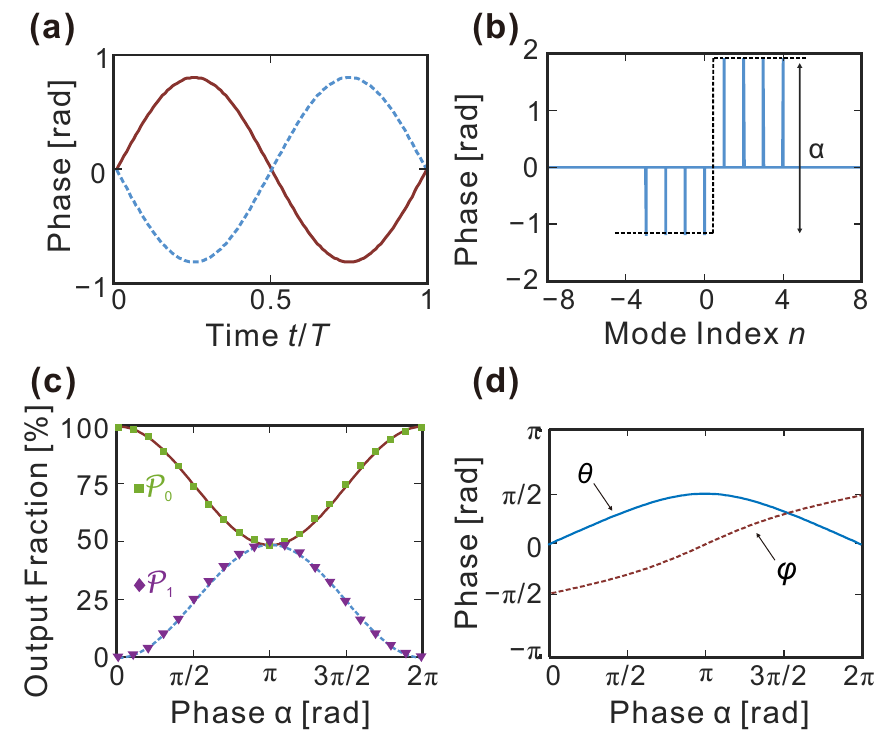}
\caption{Tunable beamsplitter. (a) Temporal phase modulation applied to the first EOM (solid red) and second EOM (dotted blue). (b) Spectral phase pattern applied by the pulse shaper, where modes 0 and 1 denote the computational space. %The reflectivity can be tuned by changing the depth of the phase jump $\alpha$.
(c) Theoretical beamsplitter transmissivity $\mathcal{T}$ (solid red) and reflectivity $\mathcal{R}$ (dotted blue). Markers denote the values measured with single-photon-level input. (d) Extracted parameters for the corresponding unitary $U(\theta,\varphi,0)$.}
\label{figS3}
\end{figure}

Figure~\ref{figS3}(a-b) depicts the specific configuration for tunable beamsplitter design. The EOMs are driven with $\pi$-phase-shifted sinewaves with a modulation index ($\Theta$) of $0.829$~rad, and the pulse shaper applies a step function with phase jump $\alpha$ between the two computational modes. This is similar to that of our previous demonstration~\cite{Lu2018b}, with only a slight difference in the EO modulation---previously, the modulation index was set at $\Theta=0.8169$ rad, which can numerically realize a fidelity $\sF_W=0.9999$ and success probability $\sP_W=0.9760$ for the Hadamard operation (when $\alpha$ is $\pi$). In this work, the theoretical fidelity for the Hadamard gate is boosted to $\sF_W=0.9999999$ after a small reduction in the success probability, $\sP_W=0.9746$. If we write out the analytical form of the $2\times 2$ transformation matrix on bins 0 and 1 as a function of the phase jump $\alpha$,
\begin{equation}
\tag{S5}
W =
\begin{pmatrix}
W_{00}(\alpha) & W_{01}(\alpha) \\
W_{10}(\alpha) & W_{11}(\alpha)
\end{pmatrix},
\end{equation}
we can derive each of the matrix elements as 
\begin{equation}
\tag{S6}
\centering
\begin{aligned}
W_{10}(\alpha) &= W_{01}(\alpha) = (1-e^{i\alpha}) \sum_{k=1}^\infty J_k(\Theta) J_{k-1}(\Theta) \\
W_{00}(\alpha) &=  J_0^2(\Theta) + (1+e^{i\alpha})\frac{1-J_0^2(\Theta)}{2} \\
W_{11}(\alpha) &=  e^{i\alpha} J_0^2(\Theta) + (1+e^{i\alpha})\frac{1-J_0^2(\Theta)}{2},
\end{aligned}
\end{equation}
where $J_k(\Theta)$ is the Bessel function of the first kind. We can define $|W_{10}|^2=|W_{01}|^2\equiv \mathcal{R}$ (i.e., mode-hopping probability, or reflectivity) and $|W_{00}|^2=|W_{11}|^2\equiv \mathcal{T}$ (i.e., probability of preserving frequency, or transmissivity). We note that when $\alpha=0$, the transformation is the identity operation as two $\pi$-phase-shifted sinewaves cancel each other out. In addition, we have $\mathcal{R}\approx \mathcal{T}$ when $\alpha=\pi$, and the elements $\{W_{00},W_{01},W_{10}\}$ are all real and positive, while $W_{11}$ is real and negative---in accord with the Hadamard operation.

Figure~\ref{figS3}(c) plots the theoretical $\mathcal{T}$ (solid red) and $\mathcal{R}$ (dotted blue) with respect to $\alpha$. On top of the two curves we also mark the (normalized) photon counts obtained in the Pauli $Z$ measurement with single-photon-level input at frequency mode $0$, which matches the theoretical prediction well. Finally, to visualize the trajectory of such a beamsplitting operation on the Bloch sphere, we map the output state $W_{00}(\alpha)\ket{0}+W_{10}(\alpha)\ket{1}$ to the form of $\cos\frac{\theta}{2}\ket{0}+e^{i\varphi}\sin\frac{\theta}{2}\ket{1}$ from the equivalent unitary $U(\theta,\varphi,0)$ and compute the corresponding $\theta$ and $\varphi$ [shown in Fig.~\ref{figS3}(d)]. The maximum $\theta$ goes to $\pi/2$ (i.e., from north pole to equator) since our tunable beamsplitter design achieves maximum reflectivity of $50\%$, rather than a full frequency hop. As for $\varphi$, it follows a continuous path from $-\pi/2$ to $\pi/2$, which explains the counterclockwise trajectory shown in Fig.~4(a) of the main text. Note that for better visualization, we wrap the parameter $\varphi$ within $-\pi$ and $\pi$ (instead of $0$ to $2\pi$ as in the main text).

\end{document}